\documentclass[12pt]{article}

\usepackage{graphicx}
\usepackage{amsfonts,amsmath,amssymb,amsthm}
\usepackage{indentfirst}
\usepackage{cite,hyperref}
\usepackage{mathptmx}

\numberwithin{equation}{section}
\DeclareMathAlphabet{\mathcal}{OMS}{cmsy}{b}{n}

\begin{document}

\title{Born-Infeld electrodynamics in very special relativity}

\author{R. Bufalo$^{1,2}$\\
\textit{{$^{1}${\small Department of Physics, University of Helsinki, P.O. Box 64}}}\\
\textit{\small FI-00014 Helsinki, Finland}\\
\textit{{$^{2}${\small Instituto de F\'{\i}sica Te\'orica (IFT), Universidade Estadual Paulista}}} \\
\textit{\small Rua Dr. Bento Teobaldo Ferraz 271, Bloco II Barra Funda, CEP
01140-070 S\~ao Paulo, SP, Brazil}\\
}
\maketitle
\date{}

\begin{abstract}
In this work we discuss the properties of a modified Born-Infeld electrodynamics
in the framework of very special relativity (VSR). This proposal allows us to
study VSR mass effects in a gauge-invariant context of nonlinear electrodynamics.
It is analyzed in detail the electrostatic solutions for two different cases, as well as
the VSR dispersion relations are found to be of a \emph{massive} particle with
nonlinear modifications. Afterwards, the field energy and static potential are computed,
in the latter we find from the VSR contribution a novel long-range $1/L^3$ correction to the Coulomb potential, in contrast to the $1/L^5$ correction of the usual Born-Infeld theory.
\end{abstract}

\newpage

\section{Introduction}

\label{sec:1}

In light of Planck scale physics, $E_{P}=\sqrt{\hbar c^5 /G}$, many theories of quantum gravity predict breaking of some symmetry groups \cite{ref4}.
In particular, the possible violation of the underlying Lorentz symmetry, one of the cornerstone of high-energy physics,
has been received many attention in several proposals which ought to incorporate such violating effects into the foundations of both general relativity
and quantum field theory. The most interesting candidates in such a quest are Standard Model Extension (SME), doubly special relativity (DSR) and very special relativity (VSR),
a common feature on the last proposals is that they enhance the Lorentz algebra by modifying the dynamics of particles.
In the SME scenario fixed background tensor fields are considered to couple with the standard model fields \cite{ref3}.
This coupling gives rise to Lorentz violating effects, parametrized by higher-order operators conceived as the vacuum expectation values of
some Lorentz fields belonging to the underlying theory.
Besides, DSR is a framework where properties of flat quantum space-time are encoded, for instance, into
(leading-order) nonlinear modifications of the energy-momentum relations $E^2 = p^2 + m^2 + \alpha l_{P} E^3 +...$ \cite{ref1}, where the leading effects are probed in $l_{P} =1/E_{P}$. 

In the VSR framework, Cohen and Glashow proposed that the laws of physics are not invariant under the whole Lorentz group $SO(1, 3)$ (with six parameters)
but rather are invariant under subgroups of the Lorentz group preserving the basic elements of special relativity \cite{ref5,ref6}, for instance, conservation laws and (unlike the DSR) the usual relativistic dispersion relation, $E^2 = p^2 + M^2$ for a particle of mass M, and etc. In particular, within this proposal, new effects are encoded in the form Lorentz-violating terms in the Lagrangian that are necessarily nonlocal and break discrete spacetime symmetries, including $CP$; but if $CP$ is incorporated as a symmetry, then the Lorentz group is again recovered. There are two subgroups fulfilling the aforementioned requirements, namely, the $HOM(2)$ (with three parameters) and the $SIM(2)$ (with four parameters) \cite{ref5}. The former is the so-called Homothety group, generated by $T_1 = K_x + J_y$,  $T_2 = K_y - J_x$, and $K_z$ ($\vec{J}$ and $\vec{K}$ are the generators of rotations and boosts, respectively). The latter, called the similitude group $SIM(2)$, is the $HOM(2)$ group added by the $J_z$ generator. A realization of VSR via a lightlike noncommutative deformation of Poincar\'e symmetry was discussed in Ref.~\cite{ref8}.

A remarkable observable consequence of VSR is a novel mechanism for introducing neutrino
masses without introducing new particles \cite{ref6}. This follows from the observation that a $SIM(2)$-covariant Dirac equation has the form
\begin{equation}
\left(i \gamma ^{\mu}\tilde{\partial}_{\mu} -M \right) \Psi \left(x\right) =0,
\end{equation}
where the wiggle operator is defined by $ \tilde{\partial}_{\mu}=\partial_{\mu}+\frac{1}{2}\frac{m^{2}}{\left(n.\partial\right)}n_{\mu} $, with a chosen preferred null direction 
$n_{\mu}=\left(1,0,0,1\right)$ that transforms multiplicatively under a VSR transformation. Thus, if we square the $SIM(2)$-modified Dirac operator, one obtains
\begin{equation}
\left[\partial_{\mu}\partial^{\mu} +\mathcal{ M}^2 \right] \Psi \left(x\right) =0, \quad \mathcal{ M}^2 = M^2 +m^2 .
\end{equation}
Hence, even with massless fields, $M=0$, the physical mass $\mathcal{ M}$ need not vanish due to VSR effects. Moreover, $m$ was introduced by dimensional reason
and we see that it sets the scale of VSR effects. VSR generalizations have been studied in the
context of nonrelativistic limit of the Dirac equation \cite{ref31} and the hydrogen atom \cite{ref40}, as well as in curved space-time \cite{ref11}, also to include supersymmetric extensions \cite{ref7} and Abelian and non-Abelian gauge fields \cite{ref9,ref14}, and to explore consequences in cosmology \cite{ref16}.

Although VSR generalizations for Maxwell and Yang-Mills field theories have been developed in Refs.~\cite{ref9,ref14}, respectively, nonlinear 
electromagnetic theories, for instance, Born-Infeld (BI) \cite{ref17} and Euler-Heisenberg \cite{ref18}, have not been considered in this framework so far.
Nonlinear extensions of Maxwell electrodynamics are well motivated both by theoretical and experimental reasons \cite{ref19,ref20},
where the field energy of a pointlike charge is finite and the phenomenon of vacuum birefringence is present in some classes of models \cite{ref32}.
In recent years a renewed interest in nonlinear electrodynamics has been raised, where several generalized nonlinear field models were proposed \cite{ref21,ref22,ref24}.
In addition, interest in nonlinear electrodynamics is also present in the context of gravitational \cite{ref28} and black hole physics \cite{ref30},
as well as in holographic thermalization \cite{ref39}.

In particular, it should be emphasized that the gauge fields in the $SIM(2)$-invariant Maxwell theory preserve the two polarization states, while they can have nonzero mass  \cite{ref9}.
This feature is remarkably interesting, since massive fields can be described by modified gauge transformations. This possibility will
be explored in full detail here in the context of Born-Infeld electrodynamics. Actually, related issues involving mass effects in nonlinear electrodynamics have received some attention,
they were studied in the context of a $\left(2+1\right)$-dim. topologically massive Born-Infeld theory \cite{ref38} and recently in a $\left(3+1\right)$-dim. massive supersymmetric Born-Infeld theory \cite{ref33}.

The letter is organized as follows. In Sec.~\ref{sec:2}, we review and present the main
aspects of VSR Abelian gauge fields, in order to derive a proper field strength. In Sec.~\ref{sec:3}, we introduce the VSR modified generalized Born-Infeld theory. 
We evaluate the electrostatic solutions, obtaining its closed form expressions as well as the leading contribution in $m$. Moreover, we obtain the (transverse and parallel polarizations) dispersion relations by considering the propagation of a electromagnetic wave, and show that these have the profile of a (nonlinear) modified \emph{massive} particle. In conclusion, we
derive the field energy and gauge-invariant potential; in the latter, we show
that the VSR contribution gives a novel long-range $1/L^3$ correction to the Coulomb potential,
in contrast to the $1/L^5$ correction of the usual Born-Infeld theory.
In Sec.~\ref{sec:4} we summarize the results, and present our final remarks.


\section{$SIM(2)$-invariant Abelian gauge fields}

\label{sec:2}

In this section we will briefly review the general aspects concerning the $SIM(2)$ modified Maxwell theory \cite{ref9,ref14}.
That will allows us to find the VSR field strength that will be used in order to construct a VSR modified Born-Infeld electrodynamics.
The first point to emphasize is that the gauge transformation of a gauge field $A_{\mu}$ in VSR is modified as, 
\begin{equation}
\delta A_{\mu}=\tilde{\partial}_{\mu}\Lambda,\label{eq: 0.1}
\end{equation}
 where the wiggle operator $\tilde{\partial}_{\mu}$ is defined as before. Now, let us consider a charged scalar field $\phi$ with gauge transformation, $\delta\phi=i\Lambda\phi $. An important step is to construct a covariant derivative that satisfies the fundamental property of transforming 
\begin{equation}
\delta\left(D_{\mu}\phi\right)=i\Lambda\left(D_{\mu}\phi\right).\label{eq: 0.4}
\end{equation} 
It can be shown that the operation defined by \cite{ref9,ref14}
\begin{equation}
D_{\mu}\phi=\partial_{\mu}\phi-iA_{\mu}\phi+\frac{i}{2}m^{2}n_{\mu}\left(\frac{1}{\left(n.\partial\right)^{2}}\left(n.A\right)\right)\phi,\label{eq: 0.3}
\end{equation}
actually is consistent with \eqref{eq: 0.4}. By this very reason we shall call the operator $D_{\mu}\phi$ as the covariant derivative
of $\phi$. In the same as for the ordinary derivative, we can define a wiggle covariant derivative by%
\begin{equation}
\tilde{D}_{\mu}\phi=D_{\mu}\phi+\frac{1}{2}\frac{m^{2}}{\left(n.D\right)}n_{\mu}\phi,\label{eq: 0.5}
\end{equation}
where it reduces to the wiggle operator $\tilde{\partial}_{\mu}$ as $A_{\mu} =0$. We are now in a position to determine the modified field strength.
In order to determine the field strength associated to $D_{\mu}$ we can consider the standard computation of the following quantity $\left[D_{\mu},D_{\nu}\right]\phi =-iF_{\mu\nu}\phi$,
where we have defined the field strength of $A$ by
\begin{equation}
F_{\mu\nu}=\partial_{\mu}A_{\nu}+\frac{m^{2}}{2}n_{\mu}\left(\frac{1}{\left(n.\partial\right)^{2}}\partial_{\nu}\left(n.A\right)\right) - \mu \leftrightarrow \nu.\label{eq: 0.6}
\end{equation}
 However, as one would naively expect, this field-strength does not coincide with the quantity 
\begin{equation}
\tilde{F}_{\mu\nu}= \tilde{\partial}_{\mu}A_{\nu}-\tilde{\partial}_{\nu}A_{\mu}.
\end{equation}
Besides, such quantity is gauge invariant and it will be used to describe massive gauge fields in the BI electrodynamics.
Nevertheless, by considering that the difference between them must be gauge invariant as well, we may then define the wiggle field strength by 
\begin{equation}
\tilde{F}_{\mu\nu} =F_{\mu\nu}+\frac{m^{2}}{2}\frac{1}{\left(n.\partial\right)^{2}} \left(n_{\nu}n^{\lambda} F_{\mu\lambda}-n_{\mu} n^{\lambda}F_{\nu\lambda}\right).\label{eq: 0.7}
\end{equation}
This may be considered our starting point. Some few comments are now in place. Using the wiggled definition of the field strength \eqref{eq: 0.7}, we can write
the simplest gauge invariant quadratic form
\begin{equation}
S=\int d^{4}x\left[-\frac{1}{4}\tilde{F}_{\mu\nu}\tilde{F}^{\mu\nu}\right],\label{eq: 0.8}
\end{equation}
this action can be added by further gauge invariant quadratic terms as well as by charged scalar or spinor fields \cite{ref14}.
By means of illustration, let us consider the equations of motion from the action \eqref{eq: 0.8},
\begin{equation}
\tilde{\partial}_{\lambda}\tilde{F}^{\lambda\mu}=0,
\end{equation}
also, by considering a VSR-type Lorenz condition, $\tilde{\partial}^{\mu}A_{\mu}=0$, we then find that
\begin{equation}
\tilde{\partial}^{2}A_{\mu}=\left(\partial^{2}+m^{2}\right)A_{\mu}=0.
\end{equation}
This shows that a Proca field, in terms of the ordinary derivative, can be described properly in a gauge-invariant fashion in terms of the wiggle operator. \footnote{One can show that the wiggle operator gives the field a mass $m$, for instance, this can be seen by the action $\int dx\phi\tilde{\partial}^{2}\phi=\int dx\phi\left(\partial^{2}-m^{2}\right)\phi$. }


\section{VSR Born-Infeld-like model }

\label{sec:3}

Let us now describe the model under consideration. In a general form, we shall consider an action profile known as generalized Born-Infeld
electrodynamics, in which we have an extended theory with two parameters \cite{ref21}. Since we have already
derived the wiggle field strength \eqref{eq: 0.7}, we can define the following nonlinear Lagrangian density 
\begin{equation}
\mathcal{L}=\beta^{2}\left[1-\left(1+\frac{2}{\beta^{2}}\mathcal{F}-\frac{1}{\beta^{2}\gamma^{2}}\mathcal{G}^{2}\right)^{p}\right], \label{eq: 1.0}
\end{equation}
where the invariants are defined as the following $\mathcal{F}=\frac{1}{4}\tilde{F}_{\mu\nu}\tilde{F}^{\mu\nu}$,
and $\mathcal{G}=\frac{1}{4}\tilde{F}_{\mu\nu}\tilde{G}^{\mu\nu}$. As usual, the dual electromagnetic field strength tensor is given
by $\tilde{G}^{\mu\nu}=\frac{1}{2}\epsilon^{\mu\nu\rho\lambda}\tilde{F}_{\rho\lambda}$.
In the Lagrangian density \eqref{eq: 1.0} the parameter $p$ is considered to dwells into the domain $0<p<1$, this reason will be clearer below.
In particular, the BI Lagrangian is given as $p=1/2$. Nonetheless, we will consider an arbitrary $p$ in the following general discussion.
The equations of motion for the VSR BI theory follow from the Lagrangian density \eqref{eq: 1.0},
\begin{equation}
\tilde{\partial}_{\lambda}\left[\frac{1}{\Pi^{1-p}}\left[\tilde{F}^{\lambda\mu}-\frac{1}{\gamma^{2}}\mathcal{G}\tilde{G}^{\lambda\mu}\right]\right]=0,\label{eq: 1.1}
\end{equation} with the quantity: $\Pi=1+\frac{2}{\beta^{2}}\mathcal{F}-\frac{1}{\beta^{2}\gamma^{2}}\mathcal{G}^{2}$. Moreover, we have that the Bianchi identities
are written as $\tilde{\partial}_{\lambda}\tilde{G}_{\mu\nu}=0$.

Now, from the temporal component of Eq.\eqref{eq: 1.1}, we find the Gauss's law
\begin{equation}
\tilde{\nabla}.\tilde{\mathbf{D}}=0,\label{eq: 1.3}
\end{equation}
where the vector field $\tilde{\mathbf{D}}$ is given by
\begin{equation}
\tilde{\mathbf{D}}=\frac{\left[\tilde{\mathbf{E}}+\frac{1}{\gamma^{2}}\left(\tilde{\mathbf{E}}.\tilde{\mathbf{B}}\right)\tilde{\mathbf{B}}\right]}{\left[1+\frac{1}{\beta^{2}}\left(\tilde{\mathbf{B}}^{2}-\tilde{\mathbf{E}}^{2}\right)-\frac{1}{\beta^{2}\gamma^{2}}\left(\tilde{\mathbf{E}}.\tilde{\mathbf{B}}\right)^{2}\right]^{1-p}}.\label{eq: 1.4}
\end{equation}
with the definitions for the electric and magnetic fields $\tilde{E}^{i}\equiv\tilde{F}^{i0}$ and $\tilde{B}^{i}=\frac{1}{2}\epsilon^{ijk}\tilde{F}_{jk}$, respectively. In particular, if we consider a current density for pointlike charge $J^{0}\left(t,\mathbf{r}\right)=g\delta^{\left(3\right)}\left(\mathbf{r}\right)$,
and by defining a new static field by $\tilde{D}_{i}\left(x\right)=\tilde{\partial}_{i}K\left(x\right)$, we have that
\begin{equation}
\left(-\nabla^{2}+m^{2}\right)K=J^{0},
\end{equation}
this equation has a solution as the Yukawa potential, yielding to
\begin{equation}
K\left(r\right)=\frac{g}{4\pi r}e^{-mr}.
\end{equation}
From this result we immediately find, $
\tilde{\mathbf{D}} = Qe^{-mr}\left(1+mr\right)\tilde{\nabla}\left(\frac{1}{r}\right)$,
where the wiggle derivative reads
\begin{equation}
\tilde{\nabla}\left(\frac{1}{r}\right)=\frac{\hat{\mathbf{r}}}{r^{2}}+\frac{m^{2}\hat{\mathbf{n}}}{2}\left[\frac{1}{\left(\hat{n}.\nabla\right)}\frac{1}{r}\right],
\end{equation}
where $\hat{\mathbf{n}}=\left(0,0,1\right)$. However, the nonlocal term of the above expression can be worked out with a distributional result as the following \cite{ref35}:
let us consider an arbitrary operator $A$, its Fourier transformation can be defined as follows
$\left(Ad\right)\left(x\right)=\frac{1}{2\pi}\int dka\left(ik\right)\hat{d}\left(k\right)e^{ikx}$, where $a\left(ik\right)$ is some function associated to the operator
$A$ \cite{ref35}. Hence, this representation can be used to define inverse-differential
operators. In particular, we have that $\frac{1}{4\pi r}=\int\frac{d^{3}k}{\left(2\pi\right)^{3}}\frac{1}{\left|\vec{k}\right|^{2}}e^{i\vec{k}.\vec{r}}$
and therefore this result can used to show that
$\left(\frac{1}{\left(\hat{n}.\nabla\right)}\frac{1}{4\pi r}\right)=\frac{1}{2\pi^{2}}\int_{0}^{\infty}\frac{dw}{w}Si\left(w\right)=\frac{1}{2\pi^{2}}\left[\frac{\gamma\pi}{2}+\underset{x\rightarrow+\infty}{\lim} \ln\left(x\right)Si\left(x\right)\right]\equiv\Lambda\left(x\right)$,
with $x\rightarrow +\infty$ and $Si\left(x\right)$ is the sine integral function. Unfortunately we see that this distribution is not regular, therefore
the cut-off $\Lambda\left(x\right)$ diverges as $x\rightarrow +\infty$.

From the above discussion we then find the solution
\begin{equation}
\tilde{\mathbf{D}}\left(\mathbf{r}\right)=Qe^{-mr}\left(1+mr\right)\left[\frac{\hat{\mathbf{r}}}{r^{2}}+\frac{m^{2}\Lambda\left(x\right)}{2}\hat{\mathbf{n}}\right]. \label{eq: 1.5}
\end{equation}
In particular, for the electrostatic case, we have the electric field, $\xi=1-p$,
\begin{equation}
\frac{\left|\tilde{\mathbf{E}}\right|}{\left[1-\frac{\tilde{\mathbf{E}}^{2}}{\beta^{2}}\right]^{\xi}}=\frac{Q e^{-mr}}{r^{2}}\left(1+mr\right)\left[1+\frac{m^{2}r^{2}\Lambda }{2}\left(\hat{\mathbf{n}}.\hat{\mathbf{r}}\right)\right]. \label{eq: 1.5a}
\end{equation}
This expression is very important in order to bound the allowed values for the parameter $p$. If we require that the electrostatic field
be regular at the origin ($r\rightarrow 0$, where it acquires its maximum
at $E_{m}=\beta$) and also that it does not blow up (for some field configurations), we find that this is achieved only when $0<p<1$.

More importantly, we can see from \eqref{eq: 1.5a} that the nonlocal VSR effects are not only present in the $\hat{\mathbf{n}}$ direction, but also in the first term
since we have the presence of $m$. Therefore, since we are interesting in the leading contributions in $m$, we shall disregard the cut-off $\Lambda\left(x\right)$ term,
and focus our study in exploring VSR deviations coming from the first term. This approximation can be seen as if we are describing a (true) massive field (Yukawa potential).

Nonetheless, taking into account the above considerations, let us consider solutions for some particular values of $p$. In particular, the BI case is given by $p=1/2$, and we obtain 
\begin{equation}
\left|\tilde{\mathbf{E}}\right|=\frac{Q e^{-mr}\left(1+mr\right)}{\sqrt{r^{4}+\frac{Q^{2}}{\beta^{2}}e^{-2mr}\left(1+mr\right)^{2}}}, \label{eq: 1.6}
\end{equation}
where, for $m=0$, we recover the usual BI solution $\left|\tilde{\mathbf{E}}\right|=\frac{Q }{\sqrt{r^{4}+ Q^{2}/ \beta^{2}}}$.
Moreover, by calculation purposes, the leading contribution in $m$ from \eqref{eq: 1.6} is found to be
\begin{align}
\left|\tilde{\mathbf{E}}\right| & =\frac{Q}{\sqrt{r^{4}+Q^{2}/\beta^{2}}}\left[1-\frac{m^{2}r^{6}}{2}\frac{1}{\left[r^{4}+Q^{2}/\beta^{2}\right]}\right].\label{eq: 1.7}
\end{align}
Besides, we can also consider another interesting case in which $p=3/4$ \cite{ref22}, \footnote{ The effective action for the $(2+1)$ QED describing a single layer graphene is driven by a power $3/4$ \cite{ref36}.} and the electrostatic field reads
\begin{equation}
\left|\tilde{\mathbf{E}}\right|=\frac{\sqrt{2}\beta \Xi }{\sqrt{\Xi^{2}+\sqrt{\Xi^{4}+4\beta^{4}}}},
\end{equation}
where we have defined $\Xi \equiv \frac{Q}{r^{2}}e^{-mr}\left(1+mr\right)$. Again, for $m=0$, we find $\left|\tilde{\mathbf{E}}\right|=\frac{\sqrt{2}\beta Q}{\sqrt{Q^{2}+\sqrt{4\beta^{4}r^{8}+Q^{4}}}}$ \cite{ref22}. Moreover, the first deviation is found as
\begin{align}
\left|\tilde{\mathbf{E}}\right|  = \frac{\sqrt{2}Q\beta}{\sqrt{Q^{2}+\sqrt{4\beta^{4}r^{8}+Q^{4}}}} \bigg[1-\frac{m^{2}r^{2}}{2} \frac{\sqrt{4\beta^{4}r^{8}+Q^{4}}}{Q^{2}+\sqrt{4\beta^{4}r^{8}+Q^{4}}}\left[1+\frac{Q^{2}}{2\beta^{2}r^{4}}\right]\bigg].\label{eq: 1.8b}
\end{align}

We shall now focus our discussion on the propagation of electromagnetic waves for the case $p=1/2$. For that matter, it is advantageous to introduce $\mathbf{\tilde{D}}=\partial\mathcal{L}/\partial\tilde{\mathbf{E}}$ and $\tilde{\mathbf{H}}=-\partial\mathcal{L}/\partial\tilde{\mathbf{B}}$,
in analogy to the electric displacement and magnetic field strength,
\begin{align}
\mathbf{\tilde{D}}&=\frac{1}{\Pi^{1/2}}\left(\tilde{\mathbf{E}}+\frac{1}{\gamma^{2}}\left(\tilde{\mathbf{E}}.\tilde{\mathbf{B}}\right)\tilde{\mathbf{B}}\right), \\
\tilde{\mathbf{H}}&=\frac{1}{\Pi^{1/2}}\left(\tilde{\mathbf{B}}-\frac{1}{\gamma^{2}}\left(\tilde{\mathbf{E}}.\tilde{\mathbf{B}}\right)\tilde{\mathbf{E}}\right),
\end{align}
 where $\Pi=1+\frac{1}{\beta^{2}}\left(\tilde{\mathbf{B}}^{2}-\tilde{\mathbf{E}}^{2}\right)-\frac{1}{\beta^{2}\gamma^{2}}\left(\tilde{\mathbf{E}}.\tilde{\mathbf{B}}\right)^{2}$.
From such constitutive relations we can introduce the tensors electric permittivity $D_{i}=\varepsilon_{ij}E_{j}$, and magnetic permeability, $B_{i}=\mu_{ij}H_{j}$. We have then that the corresponding VSR Maxwell equations are written as
\begin{gather}
\tilde{\nabla}.\tilde{\mathbf{D}}=0,\quad\tilde{\partial}_{t}\tilde{\mathbf{D}}+\tilde{\nabla}\times\tilde{\mathbf{H}}=0, \\
\tilde{\nabla}.\tilde{\mathbf{B}}=0,\quad\tilde{\partial}_{t}\tilde{\mathbf{B}}+\tilde{\nabla}\times\tilde{\mathbf{E}}=0.
\end{gather}
In order to discuss the vacuum birefringence phenomenon, let us now consider the description of a weak electromagnetic wave $\left(\mathbf{E}_{p},\mathbf{B}_{p}\right)$
propagating along the $x$ axis in the presence of a strong constant electric field \textbf{$\mathbf{E}_{0}=E_{0}e_{3}$} ($\mathbf{B}_{0}=0$).
In particular, we can picture this situation as if this background field was considered as being the contributions $O\left(m^{2}\right)$ to the Electric field
in Eq.\eqref{eq: 1.7}. In such a case, we find the following constitutive relations
\begin{align}
\tilde{D}^i &=\frac{1}{\Gamma^{1/2}}
\left(\delta ^{ij}+\frac{1}{\beta^{2}}\frac{\tilde{E}_{0}^i\tilde{E}_{0}^j}{1-\frac{1}{\beta^{2}}\tilde{\mathbf{E}}_{0}^{2}}\right)\tilde{E}_{p}^j,\\
\tilde{H}^i &=\frac{1}{\Gamma^{1/2}}
\left(\delta ^{ij}-\frac{1}{\gamma^{2}}\tilde{E}_{0}^i\tilde{E}_{0}^j\right)\tilde{B}_{p}^j.
\end{align}
with $\Gamma =1-\tilde{\mathbf{E}}_{0}^{2}/\beta^{2} $. From the plane waves decomposition, we find for the Maxwell equations 
\begin{align}
\left(\frac{k^{2}}{\tilde{\omega}^{2}}-\varepsilon_{33}\mu_{22}\right)\tilde{B}_{2}=&0,\\ 
\left(\frac{k^{2}}{\tilde{\omega}^{2}}-\varepsilon_{22}\mu_{33}\right)\tilde{B}_{3}=&0,
\end{align}
where we have defined $\tilde{\omega}^{2}=\omega^{2}-\frac{\omega m^{2}}{n.k}+\frac{m^{4}}{\left(n.k\right)^{2}}$.
In particular, if we consider the approximation $n.k = n_\mu k^\mu \sim\omega$, we can compute these equations in the following two different scenarios:
i) if $\mathbf{E}\bot\mathbf{B}$ (perpendicular polarization), we have $\tilde{B}_{3}=0$, and we find the dispersion relation
\begin{equation}
\omega_{\bot}^{2}\simeq k^{2}\left(\varepsilon_{33}\mu_{22}\right)^{-1}\left[1+\sqrt{1+\frac{2m^{2}}{k^{2}}\left(\varepsilon_{33}\mu_{22}\right)}\right]+m^{2},
\end{equation}
with $\left(\varepsilon_{33}\mu_{22}\right)^{-1}=1-\tilde{\mathbf{E}}_{0}^{2}/\beta^{2}$.
Besides, ii) if $\mathbf{E}\Vert\mathbf{B}$ (parallel polarization), we have $\tilde{B}_{2}=0$, and the dispersion relation follows
\begin{equation}
\omega_{\Vert}^{2}\simeq k^2 \left(\varepsilon_{22}\mu_{33}\right)^{-1}\left[1+\sqrt{1+\frac{2m^{2}}{k^{2}}\left(\varepsilon_{22}\mu_{33}\right)}\right]+m^{2},
\end{equation}
with $\left(\varepsilon_{22}\mu_{33}\right)^{-1}=1-\tilde{\mathbf{E}}_{0}^{2}/\gamma^{2}$. Interesting enough, we
see that these VSR modified dispersion relations display a profile of \emph{massive} particles, but with nonlinear modifications.
Furthermore, this result, the presence of the vacuum birefringence phenomenon, is in accordance with the generalized BI theory \cite{ref21}.
However, we see that for the case where $\gamma=\beta$, we have identical perpendicular and parallel polarizations, i.e. $\varepsilon_{22}\mu_{33}=\varepsilon_{33}\mu_{22}$,
implying that the phenomenon of birefringence is absent, as one would expect as in the usual BI theory.

\subsection{Interaction energy and static potential}

Since our goal is to compute the electrostatic finite energy and the gauge-invariant scalar potential it suffices to our interest to consider only
the invariant $\mathcal{F}$ term in the Lagrangian density \eqref{eq: 1.0}. Moreover, this Lagrangian can be rewritten in terms of an
auxiliary field $\chi$ (in such a way that the original theory is recovered on-shell) 
\begin{equation}
\mathcal{L}=\beta^{2}\left[1-\frac{\chi}{2}\left(1+\frac{1}{2\beta^{2}}\tilde{F}_{\mu\nu}\tilde{F}^{\mu\nu} \right)-\frac{1}{2\chi}\right]. \label{eq: 1.10}
\end{equation}
In order to implement the canonical analysis, it is convenient to rewrite conveniently the nonlocal terms present in the wiggle stress
tensor $\tilde{F}_{\mu\nu}$, Eq.\eqref{eq: 0.7}. First, we see that the invariant $\mathcal{F}=\frac{1}{4}\tilde{F}_{\mu\nu}\tilde{F}^{\mu\nu}$ can be expressed as
\begin{align}
4\mathcal{F}  =f_{\mu\nu}f^{\mu\nu}+2f_{\mu\nu}\frac{m^{2}}{\left(n.\partial\right)}n^{\mu}A^{\nu}-\frac{m^{4}}{2} \left(\frac{n^{\mu} A_{\mu}}{\left(n.\partial\right)} \right)^2,
\end{align}
where $f_{\mu\nu}=\partial_{\mu}A_{\nu}-\partial_{\nu}A_{\mu}$. Therefore, in order to handle with the nonlocal terms, we can add the constraint
$\left(n.\partial\right)\phi_{\mu}-m^{2}A_{\mu}=0$, through a Lagrange multiplier. We then find a new expression for the Lagrangian density \eqref{eq: 1.10}
\begin{align}
\mathcal{L}  =\beta^{2}-\frac{\beta^{2}}{2\chi}-\frac{\beta^{2}\chi}{2}+\Lambda^{\mu}\left(\left(n.\partial\right)\phi_{\mu}-m^{2}A_{\mu}\right)-\frac{\chi}{4}\left[f_{\mu\nu}f^{\mu\nu}+2f_{\mu\nu}n^{\mu}\phi^{\nu}-\frac{1}{2}\left(n.\phi\right)^{2}\right]. \label{eq: 1.11}
\end{align}
From this expression we can then proceed to the Hamiltonian analysis. First, we have that the canonical momenta conjugated to the gauge field $A$,
\begin{align}
\pi^{\mu}  =\frac{\partial\mathcal{L}}{\partial\dot{A}_{\mu}}=-\frac{\chi}{2}\left[2f^{0\mu}+\phi^{\mu}-n^{\mu}\phi^{0}\right], \label{eq: 1.12}
\end{align}
from this relation we immediately see that $\pi^{0}\approx0$ is a primary constraint. In addition, we find another primary constraint in
$p=\frac{\partial\mathcal{L}}{\partial\dot{\chi}}\approx0$, and also that the Lagrange multiplier $\Lambda^{\mu}$ is identified as the
canonical momentum conjugated to $\phi_{\mu}$, $p_{\phi}^{\mu}=\frac{\partial\mathcal{L}}{\partial\dot{\phi_{\mu}}}=\Lambda^{\mu}$. Moreover, from the relation \eqref{eq: 1.12} we obtain the dynamical relations, $ 2f^{0i}=-\phi^{0}n^{i}+\phi^{i}-\frac{2}{\chi}\pi^{i}$. From these results, we find that
the canonical Hamiltonian is given by
\begin{align}
&\mathcal{H}_{C}  =-\frac{1}{2\chi}\pi_{i}\pi^{i}+\frac{1}{2}\pi_{i}\left[\phi^{i}-\phi^{0}n^{i}\right]+\pi^{i}\partial_{i}A_{0}+p_{\phi}^{\mu}\left[\left(\hat{n}.\nabla\right)\phi_{\mu}+m^{2}A_{\mu}\right] \nonumber \\
&
-\beta^{2}+\frac{\beta^{2}}{2\chi}+\frac{\beta^{2}\chi}{2}  +\frac{\chi}{4}\left[f_{ij}f^{ij}+\frac{1}{2}\left[\phi_{i}-\phi_{0}n_{i}\right]^{2}+\phi_{i}n_{j}f^{ij}-\frac{1}{2}\left(n.\phi\right)^{2}\right]. 
\end{align}
Proceeding with the constraint analysis \textit{\`a la} Dirac, we should require the preservation of the constraint $\Omega_{1}\left(x\right)=\pi^{0}\approx0$, for that
\begin{equation}
\left\{ \pi^{0}\left(x\right),H_{C}\right\} =\partial_{i}\pi^{i}\left(x\right)-m^{2}p_{\phi}^{0} \left(x\right),
\end{equation}
we obtain a secondary constraint, the Gauss law $\Omega_{2}\left(x\right)=\partial_{i}\pi^{i}-m^{2}p_{\phi}^{0}\approx0$.
Now, the consistency condition of this secondary constraint gives
\begin{equation}
2 \left\{ \Omega_{2}\left(x\right),H_{C}\right\} =-m^2 \Omega_{3} \left(x\right) \approx0,
\end{equation}
where we have found a tertiary constraint  $\Omega_{3}\left(x\right)=2\partial_{i}p_{\phi}^{i}+\pi^{i}n_{i}+2\left(\hat{n}.\nabla\right)p_{\phi}^{0}\approx 0$. 
Finally, its consistency condition closes as $\left\{ \Omega_{3}\left(x\right),H_{C}\right\}   =-\Omega_{2}+\left(\hat{n}.\nabla\right)\Omega_{3}$, and therefore
no further constraints are present. At last, from the consistency condition for the constraint $p\approx0$, $\left\{ p\left(x\right),H_{C}\right\} \approx0$, we determine
the auxiliary field
\begin{align}
\chi = & \sqrt{\frac{2\beta^{2}-2\pi_{i}\pi^{i}}{2\beta^{2}+f_{ij}f^{ij}+\frac{1}{2}\left(\phi_{0}n_{i}-\phi_{i}\right)^{2}+\phi_{i}n_{j}f^{ij}-\frac{1}{2}\left(n.\phi\right)^{2} }},
\end{align}
this relation can be used to eliminate the auxiliary field $\chi$. In conclusion, we see that the set of variables $\left(A_{0},\pi^{0}\right)$
and $\left(\phi_{0},p_{\phi}^{0}\right)$ can be eliminated by setting $A_{0}=0,~~\Omega_{1}=0$ and $\phi_{0}=0, ~~ \Omega_{3}=0 $, respectively, and then evaluating the Dirac brackets.
Therefore, the dynamical generator of the canonical variables is the total Hamiltonian $H=H_{C}+\int d^{3}x v_{2} \Omega _2 \left(x\right)$,
where $v_{2}$ is an arbitrary Lagrange multiplier.

Nonetheless, in accordance with the Hamiltonian analysis, we must implement subsidiary conditions in order to fix the first class constraint $\partial_{i}\pi^{i}\approx0$.
Among the several possible choices there are the Coulomb condition $\partial_{i}A^{i}\approx0$ and the Poincar\'e condition
$\int_{C_{\xi x}}dz^{\mu}A_{\mu}\left(z\right)\equiv\int_{0}^{1}d\zeta x^{i}A_{i}\left(\zeta x\right)\approx0$ \cite{ref37},
where the contour is chosen as a spacelike straight path $z^{i}=\xi^{i}+\zeta\left(x-\xi\right)^{i}$ parametrized by $\zeta$ $\left(0\leq\zeta\leq1\right)$,
besides we can take the fixed (reference) point to be $\xi^{i}=0$, without loss of generality. In this way, one can verify that the fundamental
Dirac brackets are respectively given by
\begin{equation}
\left\{ A_{i}\left(x\right),\pi^{k}\left(y\right)\right\} _{*}=\left[\delta_{i}^{k}-\frac{\partial_{i}\partial^{k}}{\nabla^{2}}\right]\delta\left(x,y\right),
\end{equation}
 and 
\begin{equation}
\left\{ A_{i}\left(x\right),\pi^{k}\left(y\right)\right\} _{*}=\delta_{i}^{k}\delta\left(x,y\right)-\partial_{i}^{x}\int_{0}^{1}d\zeta x^{k}\delta\left(\zeta x,y\right).
\end{equation}

Now that we have concluded the Hamiltonian analysis and showed its consistency, we can proceed to evaluate the finite energy stored in the
electrostatic field of a pointlike charge. This calculation is interesting by means of comparison with the usual results of the BI electrodynamics. We have that the general stress-energy tensor is written,
\begin{equation}
T_{\nu}^{\mu}=\frac{\partial\mathcal{L}}{\partial\mathcal{F}}\tilde{F}^{\mu\sigma}\tilde{F}_{\nu\sigma}+\frac{\partial\mathcal{L}}{\partial\mathcal{G}}\tilde{F}^{\mu\sigma}\tilde{G}_{\nu\sigma}-\delta_{\nu}^{\mu}\mathcal{L}.
\end{equation}
Therefore, in the electrostatic regime and for the choice $p=1/2$ in \eqref{eq: 1.0}, we can compute the total amount of electrostatic energy, $U =\int d^3 x T_{0}^{0}$,
\begin{equation}
U =2\pi\sqrt{Q^{3}\beta}\int_{0}^{\infty} dx \sqrt{x} \left[\frac{1}{\sqrt{ 1-\Omega^{\left(1/2\right)}\left(x\right)}}-1\right].
\end{equation}
where we have made the change of variables $x=\frac{\beta r^{2}}{Q}$, and defined $\Omega^{\left(1/2\right)}\equiv\frac{\left|\tilde{\mathbf{E}}\right|^{2}}{\beta^{2}}  =\frac{e^{-2\sqrt{ax}}\left(1+\sqrt{ax}\right)^{2}}{ x^{2}+e^{-2\sqrt{ax}}\left(1+\sqrt{ax}\right)^{2} } $ and $a^{2}=\frac{Qm^{2}}{\beta}$, and
the electric field is given by Eq.\eqref{eq: 1.6}. From this we can find that the leading contributions in $m$ is given by 
\begin{equation}
U =\sqrt{\frac{g^{3}\beta}{16 \pi }}\left[a_{1}-a_{2}\left(\frac{m^{2}g}{4\pi \beta}\right)+a_{3}\left(\frac{m^{2}g}{4\pi \beta}\right)^{\frac{3}{2}}+...\right], \label{eq: 1.13}
\end{equation}
with $a_{1}=1.55$, $a_{2}=2.18$ and $a_{3}=2.96$. The first term is exactly the result obtained from BI theory, and the second is the first deviation
caused by the VSR modified theory. We can extend this analysis to the case when $p=3/4$, we find that the leading contributions have the same form as
in \eqref{eq: 1.13}, but now with $a_{1}=3.40$, $a_{2}=5.23$ and $a_{3}=7.25$.

In order to conclude our discussion, let us now proceed to compute the leading contributions for the static potential energy $V$ between pointlike sources due to the VSR terms.  This can be readily obtained by considering gauge-invariant variables. In particular, we have the gauge-invariant physical field \cite{ref41}
\begin{equation}
\mathcal{A}_{\mu}\left(x\right)=A_{\mu}\left(x\right)+\partial_{\mu}\left(-\int_{\xi}^{x}dz^{\lambda}A_{\lambda}\left(z\right)\right),\label{eq: 1.14}
\end{equation}
where, as in Poincar\'e condition, the line integral is evaluated along a spacelike path from the point $\xi$ to $x$, on a fixed slice time \cite{ref37}.
It should be emphasized that the gauge-invariant fields \eqref{eq: 1.14} are in fact physical variables, since they commute with the first-class constraint (Gauss Law).  Besides, after some manipulation we find from the expression \eqref{eq: 1.14} the scalar field
\begin{equation}
\mathcal{A}_{0}\left(t,\mathbf{r}\right)=\int_{0}^{1}d\zeta r^{i}E_{i}\left(t,\zeta\mathbf{r}\right). \label{eq: 1.15}
\end{equation}
The potential energy $V$ is usually computed by means of a Hamiltonian analysis, i.e. $\left\langle H\right\rangle _{\Omega} = \left\langle H\right\rangle _{0} +V$; in this case one have the Dirac's gauge-invariant fermion-antifermion physical state $\left| \Omega \right\rangle \equiv \left| \bar {\Psi} (\mathbf{0}) \Psi (\mathbf{L})\right\rangle$ \cite{ref37}. Instead, we may equally consider an equivalent but simple framework to compute the potential $V$ \cite{ref37}. In which, one naturally uses the gauge-invariant scalar field \eqref{eq: 1.15} in order to write the potential energy $V$; in particular, for a pair of static pointlike opposite charges, i.e. $J^0 (t,\mathbf{r})=g \left[ \delta ^{3} \left(\mathbf{r}\right) - \delta ^{3} \left(\mathbf{r}-\mathbf{L}\right)  \right]$, we have
\begin{equation}
V=g\left(\mathcal{A}_{0}\left(\mathbf{0}\right)-\mathcal{A}_{0}\left(\mathbf{L}\right)\right). \label{eq: 1.16}
\end{equation}

Since we are interested in estimating the leading contributions in $m$ that perhaps may modify the profile of the lowest-order correction in $\beta$
from the BI theory, we shall consider the simplified electric field expression \eqref{eq: 1.7}. Hence, substituting this back into \eqref{eq: 1.15}, we obtain
\begin{align}
\mathcal{A}_{0} =&-\frac{Q}{r} \int_{0}^{1}\frac{d\zeta}{\sqrt{\zeta^{4}+\frac{Q^{2}}{r^{4}\beta^{2}}}}\biggl[1-\frac{m^{2}r^{2}}{2} \frac{\zeta^{6}}{\zeta^{4}+\frac{Q^{2}}{r^{4}\beta^{2}}}\biggr].
\end{align}
These integrals can be readily evaluated, and the resulting closed form expression for the potential \eqref{eq: 1.16} is found to be,
\begin{align}
V  &=\frac{4\pi Q^{2}}{L}\biggl[\frac{1}{\sqrt{\theta}}{}_{2}F_{1}\left(\frac{1}{4},\frac{1}{2};\frac{5}{4};-\frac{1}{\theta}\right)  +\frac{m^{2}L^{2}}{40}\frac{1}{\left(1+\theta\right)^{2}\theta }\biggl\{\sqrt{1+\theta}\left(10+25\theta+36\theta^2+9\theta ^3\right) \nonumber \\
 & +\sqrt{\theta}\left(10+45\theta+86\theta^2+39\theta^3\right){}_{2}F_{1}\left(-\frac{1}{2},-\frac{1}{4};\frac{3}{4};-\frac{1}{\theta}\right) \nonumber  \\
 & -4\sqrt{\theta} \left(5+\theta\left(3+2\theta\right)\left(5+6\theta\right)\right){}_{2}F_{1}\left(-\frac{1}{4},\frac{1}{2};\frac{3}{4};-\frac{1}{\theta}\right)\biggr\}\biggr], \label{eq: 1.17}
\end{align}
with $\theta=\frac{Q^{2}}{L^{4}\beta^{2}}$ and $L\equiv\left|\mathbf{L}\right|$, and where $_{2}F_{1}\left(a,b;c;z\right)$ is the hypergeometric function. Besides, we have that $\mathcal{A}_{0}\left(\mathbf{0}\right)=0$, since $_{2}F_{1}\left(a,b;c;0\right)=1$. Finally, in order to explore the behaviour of the leading corrections in the potential \eqref{eq: 1.17}, we recall the asymptotic expansion, as $z\rightarrow\infty$ and $a-b\notin\mathbb{Z}$, $_{2}F_{1}\left(a,b;c;z\right)\propto\frac{\Gamma\left(b-a\right)\Gamma\left(c\right)}{\Gamma\left(b\right)\Gamma\left(c-a\right)}\left(-z\right)^{-a}\left(1+\frac{a\left(1+a-c\right)}{\left(1+a-b\right)z}+\frac{a\left(1+a\right)\left(1+a-c\right)}{2\left(1+a-b\right)\left(2+a-b\right)z^{2}}+...\right)+a\leftrightarrow b$. Hence, employing such expansion in Eq.\eqref{eq: 1.17}, we find that the leading corrections are given by,
\begin{align}
V  =-\frac{g^{2}}{4\pi L}\biggl(1+\frac{m^{2}g^{2}}{64 \pi ^2\beta^{2}}\frac{1}{L^{2}}-\frac{g^{2}}{160 \pi ^2 \beta^{2}}\frac{1}{L^{4}} \biggr)  +\frac{g^{2}}{16 \pi^{\frac{3}{2}} }\Gamma^{2}\left(\frac{1}{4}\right) \sqrt{\frac{\beta}{Q}} \biggl(1+\frac{3\Gamma^{2}\left(\frac{3}{4}\right)}
 {4 \pi \Gamma^{2}\left(\frac{1}{4}\right)}\frac{g m^{2}}{\beta} \biggr). \label{eq: 1.18}
\end{align}
There are two important points to consider in the expression \eqref{eq: 1.18}. First, we should note the qualitative departure from the usual Maxwell theory; 
but, remarkably, the VSR contribution gives already a modified long-range $1/L^3$ correction in addition to the $1/L^5$ correction coming from the Born-Infeld theory.
Furthermore, the behaviour of this VSR modified correction is dominant in face of the usual BI correction. A subtle point, now concerning
the last two terms in \eqref{eq: 1.18}, is that the usual subtracted term in the BI theory is also corrected by a VSR contribution.


\section{Concluding remarks}

\label{sec:4}

In this paper we have analyzed the properties of a Very Special Relativity (VSR) modified
Born-Infeld electromagnetic theory. In particular, we were interested in determining
(massive) VSR deviations from the well-known electrostatic solutions from the BI theory.
We started by reviewing important features of a Abelian gauge field
defined in the VSR framework, this allowed us to obtain the respective wiggle field strength. 
By means of illustration, we showed that the gauge fields preserve the two polarization
states, while they can have nonzero mass. This particular possibility is very appealing
from the point of view of the BI theory, since only in the $(2+1)$-dim. topological
theory it is possible to describe massive modes in a gauge-invariant fashion. 

We have considered along our analysis two particular values of the generalized VSR Born-Infeld theory, $p=1/2$ and $p=3/4$. For such values, we found closed expressions for the VSR modified electric field, as well as deviations as leading contributions in $m$. Also, we have discussed the vacuum birefringence phenomenon. The resulting dispersion relations displayed a profile of a \emph{massive} particle, but with nonlinear modifications.

In conclusion, we performed an Hamiltonian analysis, computing the electrostatic finite energy and the gauge-invariant potential. Concerning the finite energy, we evaluated the
leading contributions in $m$, for the cases $p=1/2$ and $p=3/4$, where we have obtained
corrections for the ordinary value. Finally, we have computed a closed form expression
for the potential for a pair of static pointlike charges. Additionally, we have
explored the behaviour of the leading contributions, and found that the VSR
contribution induces an improved and dominant a novel long-range $1/L^3 $ correction
to the Coulomb potential, in face of the $1/L^5 $ correction from the Born-Infeld theory.


\subsection*{Acknowledgments}

R.B. thanks FAPESP for full support, Project No. 2013/26571-4.


\end{document}